\documentclass[12pt]{article}
\usepackage{epsfig,graphicx,psfrag}
\usepackage{slashed} 
\usepackage[dvips]{color}
\usepackage{color}
\definecolor{red}{rgb}{1,0,0}
\definecolor{green}{rgb}{0,1,0}
\newcommand{\be}{\begin{equation}}
\newcommand{\ee}{\end{equation}}
\newcommand{\ba}{\begin{eqnarray}}
\newcommand{\ea}{\end{eqnarray}}
\newcommand{\nn}{\nonumber}

\newcommand{\MeV}{{\rm MeV}}
\newcommand{\GeV}{{\rm GeV}}
\newcommand{\ice}[1]{\relax}

\newcommand{\MSbar}{\overline{\rm MS}}

\usepackage{color}
\usepackage{amsmath}

\usepackage{booktabs} %fancy tables
\usepackage{graphicx, subfig}
\usepackage{subfig}
\usepackage{multirow}
\begin{document}
\thispagestyle{empty}

\begin{flushright}
SI-HEP-2013-02\\
QFET-2013-01
\end{flushright}
\phantom{}
\vspace{.2cm}

\begin{center}

{\Large\bf  Decay constants of heavy-light vector mesons \\from QCD sum rules}

\vspace{1cm}
{\bf
P.~Gelhausen, A.~Khodjamirian, A.A.~Pivovarov and D.~Rosenthal
}

\vspace{2mm}
{\it  
Theoretische Physik 1, 
Naturwissenschaftlich-Technische Fakult\"at,\\
Universit\"at Siegen, D-57068 Siegen, Germany }\\[2mm]

%Institute for Nuclear Research of the\\
%Russian Academy of Sciences, 117312 Moscow, Russia

\begin{abstract}
\noindent
We revisit QCD sum rules for the 
decay constants of heavy-light mesons.
In the sum rules for the vector  mesons 
$B^*_{(s)}$ and $D^*_{(s)}$ we improve the accuracy of OPE, 
taking into account the $O(\alpha^2_s)$  
terms in the perturbative   part and calculating the $O(\alpha_s)$ corrections to 
the quark-condensate 
contribution.  With this accuracy, we obtain the ratios of decay constants:     
$f_{B^*}/f_B=1.02^{+0.02}_{-0.09}$,~ $f_{D^*}/f_D=1.20^{+0.13}_{-0.07}$. 
The sum rule predictions for the decay constants 
of pseudoscalar mesons are updated with the results $ f_B=(207^{+17}_{-9})\,\MeV$,
 $f_{B_s}=(242^{+17}_{-12})\,\MeV$, $f_D=(201^{+12}_{-13})\,\MeV$,
  $f_{D_s}=(238^{+13}_{-23})\,\MeV$.
In order to assess the sensitivity of our calculation 
to the form of the sum rule, we    
consider alternative versions such as the power moments 
and Borel sum rules with different weights of the spectral density. We also investigated
the heavy quark limit of the sum rules 
for vector and pseudoscalar mesons, estimating  the violations   
of the heavy-quark spin and flavour symmetry.
\end{abstract}

\end{center}

\newpage

\section{Introduction}
The decay constants of heavy-light mesons are the simplest hadronic matrix elements
relevant for heavy-flavour physics. The constants of 
pseudoscalar mesons, $f_{D_{(s)}}$  and $f_{B}$, 
received much attention because they determine the long-distance 
QCD dynamics in  the leptonic weak decays 
$D_{(s)}\to \mu\nu_{\mu}$ and $B\to \tau \nu_\tau$, respectively (see, e.g. 
the review \cite{RosnStone} and a more recent measurement \cite{BelleBtaunu}).
The accuracy of  $f_{B_{s(d)}}$ is vital for the analysis of 
the leptonic FCNC decay $B_{s(d)}\to \mu^+\mu^-$ \cite{LHCbBsmumu}. 
Recent determinations of the decay constants 
of $B$ and $D$ mesons in lattice QCD with dynamical flavours 
quote an impressive precision \cite{Fermilab,HPQCDc,HPQCDb,latticeav}. 

The decay constants $f_{B^*}$ and $f_{D^*}$  of the heavy-light 
vector mesons $B^*$ and $D^*$ 
cannot be directly probed in weak decays. 
Nonetheless, they also play an important role in heavy flavour phenomenology. 
To bring only one example: an accurate knowledge 
of  $f_{B^*}$ and $f_{D^*}$ is needed in the calculation of the 
strong couplings $B^*B\pi$ and $D^*D\pi$  from QCD light-cone sum rules  
\cite{Belyaevetal}.    
Furthermore, the deviation of  $f_{B^*(D^*)}$ from $f_{B(D)}$ calculated at finite masses
allows one to assess the violation of the heavy-quark spin symmetry. 
In lattice QCD, we are aware of only one recent calculation of   
$f_{D^*}$  and $f_{D^*_s}$ in \cite{fDstarlat}.

In continuum QCD, the  hadronic decay constants are determined from 
QCD (SVZ) sum rules~\cite{SVZ} based on the operator-product expansion (OPE) 
for the two-point correlation functions 
of quark currents.  The sum rules for heavy-light mesons have 
a long history, starting  from the very early papers ~\cite{NSVZ,AE}
(see also, e.g. \cite{oldSR,bounds})  in the framework of full QCD, as well as  using 
the heavy-quark expansion \cite{Shuryak},
followed by sum rules in  heavy-quark effective theory (HQET)
\cite{Neubert91,Braunetal,BroadGrozin,Neubert_rev}.  
The gluon radiative  corrections to the correlation functions at  $O(\alpha_s)$ 
(two-loop) level calculated in \cite{Broadhurst,Generalis} play an  important role.  
The sum rule for the heavy-light vector mesons with this accuracy 
can be found, e.g. in \cite{Dominguez}.

A substantial improvement of the OPE was achieved in \cite{ChetS}
where the $O(\alpha_s^2)$ (three-loop) contribution 
to the perturbative part of the correlation function was calculated. The results are available 
in a semi-numerical form. These NNLO corrections
were implemented in  the sum rule determination of the 
$B_{(s)}$ decay constants  for finite heavy quark masses in \cite{JL}, and 
in the framework of HQET in \cite{PS}. 
Note that the $\MSbar$-scheme chosen for the heavy quark mass in~\cite{JL} 
leads to a reasonable suppression of gluon radiative corrections 
in the perturbative part of the sum rules.

In this paper we revisit QCD sum rules for heavy-light meson decay constants.
Our main goal is to upgrade the accuracy of the sum rules
for the vector-meson decay constants $f_{B^*}$ and $f_{D^*}$,  including 
the three-loop,  $O(\alpha^2_s)$  terms in the OPE for the correlation function and 
calculating the missing $O(\alpha_s)$  corrections to the quark condensate contribution. 
In parallel, we update the decay constants 
of pseudoscalar heavy-light mesons. We also evaluate the 
ratios of decay constants, sensitive to the violation of 
the heavy-quark and $ SU(3)_{fl}$ symmetries in bottom and charmed mesons.
In addition to the  standard sum rule calculation, 
we obtain upper bounds for decay constants following from the positivity
of the spectral density in the dispersion relations 
and independent of the quark-hadron duality assumption. 
Furthermore, to assess 
the sensitivity of our results to the particular form of 
the sum rule with Borel exponent, we employ certain modifications of  
sum rules with different weights in the dispersion integrals, as well as  their power moments. 
 
In what follows, in Sect.~2 we present a brief outline of the method. Sect.~3
contains a discussion of the input parameters and the numerical results. Sect.~4 is devoted 
to the alternative sum rules and contains their comparison with the
conventional  Borel sum rules. In Sect.~5 we consider the heavy-quark limit of the sum rules. Our concluding discussion is presented in Sect.~6.

\section{Outline of the sum rule derivation}

Here we remind the main steps leading to the   
standard  QCD sum rule \cite{SVZ} for a decay constant. 
For the heavy-light vector mesons, which are the main objects of our study,  one
needs the following two-point  correlation function:
\begin{align}
\Pi_{\mu\nu}(q)=i\int d^4x\, e^{iqx} \langle 0| T \{j_\mu(x) j^\dagger_\nu(0)\}|0 \rangle=
\Big(\!- g_{\mu\nu}q^2+q_\mu q_\nu\Big)\widetilde{\Pi}_T(q^2)+
q_\mu q_\nu\Pi_L(q^2)\, ,
\label{eq:corrV}
\end{align}
where  $j_\mu=\bar{q}\gamma_\mu Q $ is the interpolating heavy-light quark current,  $Q=c,b$ and  $q=u,d,s$.
In this paper finite
quark masses in $\MSbar$ scheme are considered.   
In (\ref{eq:corrV}),  the relevant invariant amplitude is $\widetilde{\Pi}_T(q^2)$,  multiplying the transverse
kinematic structure. To avoid kinematical  singularities, it is more convenient to 
rescale it, introducing
\be
\Pi_T(q^2)\equiv q^2\widetilde{\Pi}_T(q^2)\,,
\label{eq:PiT}
\ee
that is, to consider  the coefficient at $-g_{\mu\nu}$ in $\Pi_{\mu\nu}(q)$.
In what follows, we also need the standard definition of 
the  correlation function with pseudoscalar 
heavy-light currents:
\be
\Pi_5(q)=i\int d^4 x \, e^{iqx} \langle 0| T \{ j_5(x) j^\dagger_5(0) \} |0 \rangle \,,
\label{eq:corr5}
\ee
where  $j_5=(m_Q+m_q)\bar{q}\,i\gamma_5 Q $.

The decay constants of the heavy-light  vector 
meson $H^*=\{B^*,D^*\}$  and pseudoscalar  meson $H=\{ B,D\} $ 
are defined in a standard way, 
\be
\langle 0|j_\mu|H^*(q) \rangle = m_{H^*}\epsilon^{(H^*)}_\mu f_{H^*} , ~~
\langle 0|j_5|H(q) \rangle = m_{H}^2 f_{H} ,
\label{eq:fBst}
\ee
where $\epsilon^{(H^*)}_\mu $ is the polarization vector of $H^*$.
These decay constants squared enter the ground-state pole terms of the 
hadronic dispersion
relations for the correlation functions (\ref{eq:corrV}) and  (\ref{eq:corr5}).  
The invariant amplitudes defined according to (\ref{eq:PiT}) and    (\ref{eq:corr5}),
obey double-subtracted dispersion relations in the variable $q^2$:
\be
\Pi_{T(5)}(q^2)  -\Pi_{T(5)}(0)-q^2\left(\frac{d\,\Pi_{T(5)}(q^2)}{dq^2}\right)\Big|_{q^2=0}=\frac{(q^2)^2}{\pi}\int ds \frac{\mbox{Im} \Pi_{T(5)}(s)}{s^2(s-q^2)}\,.
\label{eq:dispPiT}
\ee
The spectral densities 
\be 
\rho_{T} (s)\equiv \frac{1}{\pi}\mbox{Im} \Pi_{T}(s)=
m_{H^*}^2 f_{H^*}^2\delta(s-m_{H^*}^2) + \rho^h_T(s)\theta(s-(m_H+m_P)^2)\,,
\label{eq:rhoHstar}
\ee
\be
\rho_5(s)\equiv \frac{1}{\pi}\mbox{Im} \Pi_{5}(q)=m_H^4 f_{H}^2\delta(s-m_H^2)+ \rho^h_5(s)\theta(s-(m_{H^*}+m_P)^2)\,,
\label{eq:rhoH}
\ee 
are positive definite and, according to unitarity, 
are given by the sums over all hadronic contributions with the 
quantum numbers of $H^{*}$ and $H$, respectively.
In the above, the ground-state contribution  is written separately and a 
generic notation  $\rho^h_{T(5)}(s)$ is introduced for the spectral density 
of excited  and continuum states;
$m_P$ is the mass of the lightest pseudoscalar meson, $\pi$ or $K$,  depending on
 the flavour content of $H^{(*)}$. Note that in the pseudoscalar channel 
there is a gap between the ground state $H$ and the threshold $m_{H^*}+m_P$ 
of the lowest continuum state, whereas in the vector $D^*$($B^*$)  channel the 
threshold $m_H+m_P$ lies below (above but very 
close to) the ground state $H^*$. Note also that the vector mesons 
with strangeness  $D^*_s$ and $B^*_s$ 
are strongly coupled to $DK$  and $BK$ continuum states,
respectively, both having larger thresholds  than the vector meson masses, 
whereas the channels $D_s\pi$ and $B_s\pi$  are decoupled
in the limit of isospin symmetry. 

In QCD, the local OPE for the correlation function  (\ref{eq:corrV})  is valid 
at $q^2\ll m_Q^2$ , far from hadronic thresholds. 
This expansion includes a perturbative part and contributions 
of the vacuum condensates. The latter, ordered according to their operator 
dimension $d$, are taken into account up to $d=6$:
\be
\Pi^{OPE}_{T(5)}(q^2)=\Pi_{T(5)}^{(pert)}(q^2)+
\Pi_{T(5)}^{\langle \bar{q} q \rangle} (q^2)+
\Pi_{T(5)}^{\langle GG \rangle} (q^2)+
\Pi_{T(5)}^{\langle \bar{q} G q \rangle} (q^2)+
\Pi_{T(5)}^{\langle \bar{q} q \bar{q} q\rangle} (q^2)\,,
\label{eq:OPE}
\ee
where the contributions of the  quark, gluon, quark-gluon, and four-quark condensates
with dimensions $d=3,4,5,6$ are indicated by  the indices 
$ \langle \bar{q} q \rangle $, $\langle GG \rangle$, $ \langle \bar{q}G q \rangle$ and $\langle \bar{q} q \bar{q} q\rangle$,  respectively.
The contributions of $d=4,5,6$ condensates are very small in the region 
where we consider OPE, hence it is justified to neglect the 
terms of OPE stemming from the condensates  of larger dimension.
The perturbative part of the OPE is usually represented in a form of 
a dispersion integral:
\be
\overline{\Pi}_{T(5)}^{(pert)}(q^2)= (q^2)^2\!\!\!\!\!\!\int \limits_{(m_Q+m_q)^2}^\infty\!
\!\!\! ds \,
\frac{\rho_{T(5)}^{(pert)}(s)}{s^2(s-q^2)}\,.
\label{eq:pert}
\ee
Here, as in (\ref{eq:dispPiT}), the two subtractions are needed due to the $s\to \infty$  asymptotics
of the leading order (LO) perturbative contributions:
\be
\rho_{T}^{(pert,LO)}(s)=\frac{1}{8\pi^2}s\left(1-z\right)^2
\left(2 +z\right)\,,
\label{eq:pertLOT}
\ee
\be
\rho_{5}^{(pert,LO)}(s)=\frac{3(m_Q+m_q)^2}{8\pi^2}s\left(1-z\right)^2,
\label{eq:pertLO5}
\ee
given by the simple heavy-light loop diagrams, where $z=\frac{m_Q^2}{s}$. 
In the above, for simplicity, the 
light-quark masses are neglected, except in the factor $(m_Q+m_q)$ for the pseudoscalar-current density. The relevant mass corrections are presented in  Appendix \ref{App:perturbative}.

The perturbative part of OPE includes  
NLO (two-loop)  and NNLO (three-loop) terms:
\be
\rho_{T(5)}^{(pert)}(s)= \rho_{T(5)}^{(pert,LO)}(s)+
\left(\frac{\alpha_s}{\pi}\right)\rho_{T(5)}^{(pert,NLO)}(s)+
\left(\frac{\alpha_s}{\pi}\right)^2\rho_{T(5)}^{(pert,NNLO)}(s)\,.
\label{eq:rhopert}
\ee
The NLO corrections are known from early papers \cite{Broadhurst,Generalis}
and are determined by a sum of three two-loop diagrams 
originating from gluon-exchanges in the heavy-light quark loop.
We have once more recalculated these diagrams 
for vector- and pseudoscalar-current 
correlation functions  in both pole- and $\MSbar$- mass (our default) 
scheme and confirmed the results  of previous calculations \cite{Broadhurst,Generalis},
given for pseudoscalar currents, e.g. in \cite{JL}  and for vector currents 
in \cite{Dominguez}. The analytical formulae for the NLO gluon radiative corrections 
are presented in the Appendix \ref{App:perturbative}.

The NNLO corrections were calculated in \cite{ChetS} 
and implemented in the sum rule determination of  
$f_{B_{(s)}}$ in \cite{JL} and $f_{B,D}$ in \cite{PS}.  In this paper,
we include these corrections also in the 
OPE of the vector-current correlation function. 
To this end, we follow the $\alpha_s$ expansion of the invariant amplitude 
$\Pi^v(q^2)$ introduced in \cite{ChetS}, which   in our notation corresponds  to 
$\widetilde{\Pi}_T(q^2)=\Pi_T(q^2)/q^2$. For the 
three-loop, $O(\alpha_s^2)$ term in this expansion we make use of the colour-structure decomposition  
given in eq.(8) of \cite{ChetS}. For the imaginary parts 
of separate contributions  we employ the formulae obtained from the semi-numerical Pade procedure
and encoded in the program {\it Rvs.m} made available by the
authors of ref.~\cite{ChetS}.

In the OPE  (\ref{eq:OPE})  the contribution of the 
 $d=3$ quark condensate includes LO and NLO 
(one-loop, $O(\alpha_s)$) 
contributions. The latter for the pseudoscalar-current correlation function 
were calculated in \cite{Neubert91,JL}. We repeated this calculation 
and confirm their result.
The NLO correction to the quark-condensate term in the vector-current correlation function is new. Finally, for the $d=4,5,6$  contributions  of
gluon, quark-gluon and four-quark condensates to OPE,  
we use the LO expressions known from  the literature \cite{SVZ,AE,Dominguez}. 
For completeness, these expressions are also collected in the Appendix \ref{App:nonperturbative}. 

In the case of nonstrange heavy-light mesons, the $u,d$ quark 
masses are neglected everywhere in the correlation function, being 
negligibly small numerically in comparison with all other energy-momentum scales. 
In the sum rules for the strange heavy-light mesons the $s$ quark mass is taken into account 
in the prefactor $(m_Q+m_s)^2$ of $\Pi_5$. Apart from that, we include
 the $O(m_s^2)$ terms  in the LO  ($O(m_s)$ in the NLO)  
perturbative part and the $O(m_s)$ terms in the 
LO  quark-condensate contribution in both correlation functions 
(see the expressions in Appendix \ref{App:perturbative}).

Substituting the OPE result (\ref{eq:OPE}) in the hadronic dispersion relation (\ref{eq:dispPiT}),
we perform the standard Borel transformation, after which 
the subtraction terms vanish. One has: 
\be
 \Pi^{OPE}_{T}(M^2)=  m_{H^*}^2 f_{H^*}^2e^{-m^2_{H^*}/M^2}+
\!\!\!\!\!\!\int\limits_{(m_H+m_P)^2}^\infty \!\!\!\!\! ds\,\, \rho^h_T(s)e^{-s/M^2}
\label{eq:Borel}
\ee 
and a similar  relation for the pseudoscalar meson case. 
The truncated OPE on  l.h.s. is reliable at $M^2>M_{min}^2$, where the  
$d=4,5,6$ condensate contributions remain
sufficiently small numerically.  On the other hand, 
below a certain  upper boundary $M^2<M_{max}^2$ 
the contributions of 
higher states accumulated in the integral on r.h.s. in (\ref{eq:Borel})
are suppressed with respect to the ground-state term due to the Borel exponent. 
As usual, we apply the quark-hadron duality 
approximation for the hadronic spectral density $\rho^h_T(s)$
and replace the integral on the r.h.s. of (\ref{eq:Borel}) by the integral over 
the perturbative spectral density, introducing an effective threshold:
\be
\rho^h_T(s)\theta(s-(m_H+m_P)^2)= \rho^{(pert)}_T (s) \theta(s-s^{H^*}_0)\,.
\label{eq:dual}
\ee

After subtracting the part
of the dispersion integral over the perturbative spectral density from both sides 
of (\ref{eq:Borel}),  
the final form of QCD sum rule reads:
\ba
f_{H^*}^2= \frac{e^{m_{H^*}^2/M^2}}{m_{H^*}^2 }\Bigg\{
\Pi_{T}^{(pert)}(M^2,s_0^{H^*})+
\Pi_{T}^{\langle \bar{q} q \rangle} (M^2)+
\Pi_{T}^{(d456)} (M^2)
\Bigg \}\,.
\label{eq:fBstarSR}
\ea
The analogous sum rule for the pseudoscalar meson channel is: 
\ba
f_{H}^2=\frac{e^{m_{H}^2/M^2}}{m_{H}^4 }
\Bigg\{\Pi_{5}^{(pert)}(M^2,s_0^{H})+
\Pi_{5}^{\langle \bar{q} q \rangle} (M^2)+
\Pi_{5}^{(d456)} (M^2)
\Bigg \}\,.
\label{eq:fBSR}
\ea
In above equations the notation
\be
\Pi_{T(5)}^{(pert)}(M^2,s_0)=\!\!\!\!\!\!\!\int\limits_{(m_Q+m_q)^2}^{s_0}
\!\!\!\!\!\!\!ds \, e^{-s/M^2}\rho^{(pert)}_{T(5)} (s) \,,
\label{eq:Pipert}
\ee
\be
\Pi_{T(5)}^{(d456)} (M^2)=\Pi_{T(5)}^{\langle GG \rangle} (M^2)+
\Pi_{T(5)}^{\langle \bar{q} G q \rangle} (M^2)+
\Pi_{T(5)}^{\langle \bar{q} q \bar{q} q\rangle} (M^2)\,,
\label{eq:Pid456}
\ee
is used for brevity. In (\ref{eq:Pipert}) the $\alpha_s$ expansion (\ref{eq:rhopert}) of the perturbative spectral density is then substituted.
In terms of (\ref{eq:Pipert}), 
the sum over excited states and continuum contributions is 
simply equal to $\Pi_{T(5)}^{(pert)}(M^2,\infty)- \Pi_{T(5)}^{(pert)}(M^2,s_0)$.

A specific ``Borel window'' is usually adopted for each sum rule, defined 
as an interval $M_{min}^2<M^2< M_{max}^2$ where 
the higher-dimensional condensate terms in OPE and  
the excited and continuum contributions 
are suppressed simultaneously. The actual size of this interval 
depends  on the quantum numbers of quark currents in 
the correlation function.
The effective threshold 
is usually fixed by fitting the mass of the ground-state meson calculated from the sum rule to its experimentally measured value.
To obtain an equation for the meson mass squared $m^2_{H^*}$ ($m^2_{H}$) from the sum rule  
(\ref{eq:fBstarSR}) ((\ref{eq:fBSR})), it is sufficient to  differentiate 
the r.h.s of each sum rule over $(-1/M^2)$ and formally equate the result to zero.

Apart from the sum rule, the dispersion relation  
for the invariant amplitude $\Pi^{QCD}_{T(5)}$,
after Borel transformation, yields an upper bound for the decay 
constant which simply follows from the positivity 
of the hadronic spectral density and is independent of the quark-hadron duality
approximation (for earlier uses of the  
bounds see, e.g. \cite{bounds,AKbounds}).
Formally, the upper bounds for $f^2_{H^{(*)}}$ are obtained,
putting $s^{H^{(*)}}_0\to \infty$ in (\ref{eq:fBstarSR})  and (\ref{eq:fBSR}).

\section{Numerical analysis of Borel sum rules}
%%%%%%%%%%%%%%%

\begin{table}[t]
\begin{center}
\begin{tabular}{|c|c|c|}
\hline
 Parameters & Values (comments)& \\
\hline
& $\overline{m}_b(\overline{m}_b)=4.18\pm 0.03\,$GeV    &\\
quark masses 	 &  $\overline{m}_c(\overline{m}_c)=1.275\pm0.025\,$GeV
&\cite{PDG}\\
& $\overline{m}_s(2\,\GeV)=  95 \pm 10 $ MeV  (error doubled) &\\
\hline
& $\alpha_s(M_Z)=0.1184\pm 0.0007$ 	& \\
strong coupling & $\alpha_s(3 \,\GeV )= 0.255 \pm 0.003 $&\cite{PDG} \\
& $\alpha_s(1.5\,\GeV )=0.353\pm 0.006$& \\
\hline
quark condensate & $\langle\bar{q}q\rangle(2 \GeV) =-(277^{+12}_{-10}\,\text{MeV})^3 $ (ChPT $\oplus~ m_s$)&\cite{Leutwyler,PDG}\\
\hline
&$\langle\bar{s}s\rangle/\langle\bar{q}q\rangle=0.8\pm0.3$&\\
&$\langle GG\rangle= 0.012^{+0.006}_{-0.012}~\GeV^4$ &\\
$d=4,5,6$ condensates  &$m_0^2=0.8\pm 0.2\,$GeV$^2$&\cite{Ioffe} \\
& $\langle\bar{s}Gs\rangle/\langle\bar{q}Gq\rangle=\langle\bar{s}s\rangle/\langle\bar{q}q\rangle$&\\
& $r_{vac}=0.1-1.0 $ &\\
\hline
\end{tabular}
\end{center}
\caption{\it Input parameters used in the sum rules.}
\label{tab:inp}
\end{table}
%%%%%%%%%%%%%%%%
In Table~\ref{tab:inp}  the adopted intervals of  
QCD parameters  entering the sum rules
(\ref{eq:fBstarSR}) and (\ref{eq:fBSR}) are collected.  As well known, 
these sum rules are very sensitive to the value of the heavy quark mass. 
For a correlation function  of highly virtual quarks in full QCD involved in our 
calculation, we use, as in \cite{JL},  
the quark masses in the $\MSbar$ scheme. 
Currently,  there is a  good agreement between various  lattice-  and continuum-QCD
determinations of $b$ and $c$ quark masses; hence we simply take the
intervals of their  $\MSbar$ values  from~\cite{PDG}. In particular, 
the heavy quark masses extracted  from the QCD quarkonium sum 
rules~\cite{ChetmQ,Hoangmc} are very close to the average values we are using.
For the strange quark mass we  
double the theoretical uncertainty  quoted in \cite{PDG},
having in mind that the latter uncertainty is dominated by the most
recent lattice determinations. With our choice, 
the interval of continuum QCD determinations of $m_s$
(e.g. from the QCD sum rules~\cite{ms}) 
with typically larger errors, is also covered.  
The nonstrange quark condensate density
$\langle 0 |\bar{q} q |0 \rangle \equiv \langle \bar{q}q\rangle $, ($q=u,d$),
is calculated from the ChPT relations derived in \cite{Leutwyler}, using
the $s$ quark mass as an input.  The details can be found, e.g., in \cite{KKMO}.
Note that the $SU(3)_{fl}$  symmetry violation in OPE originates 
not only from the quark mass difference $m_{s}-m_{u,d}$, 
but also from the difference between
strange and nonstrange quark-condensate densities.   
For their ratio  we adopt 
a rather broad interval from the review \cite{Ioffe}, 
where also  the $d=4,5,6$ condensate densities are taken from.  
The latter  
are parametrized in a standard way: 
$
(\alpha_s/\pi)\langle 0| G^a_{\mu\nu} G^{a\,\mu\nu} |0\rangle \equiv\langle GG\rangle  $, 
$ \langle 0 |g_s\bar{q} G^a_{\mu\nu}t^a \sigma^{\mu\nu}q|0 \rangle \equiv 
\langle \bar{q}Gq\rangle =
m_0^2 \langle \bar{q}q\rangle $. Furthermore,
following \cite{SVZ}, the four-quark condensate density is factorized with an intermediate 
vacuum insertion into the square of quark condensates 
$r_{vac} \langle \bar{q}q\rangle ^2$, with  
an additional coefficient  $r_{vac}$, parameterizing the deviation from the factorization.

For the running of the QCD coupling and quark masses 
with an appropriate loop accuracy we employ 
the numerical  program {\it RunDec} available from \cite{RunDec}. 
We adopt a uniform renormalization scale $\mu$ for the correlation function,
 strong coupling and quark masses. The running of the quark-condensate
density is taken into account in the same approximation as for the quark
masses; the gluon-condensate density is renorminvariant  and the running of 
quark-gluon and four-quark condensates is negligible, i.e. their densities
$m_0^2\langle\bar{q}q\rangle$ and $\alpha_sr_{vac}\langle\bar{q}q\rangle^2$,
respectively, are taken at a low scale $\mu=1 \,\GeV$.

The choice of  the three remaining parameters 
-- Borel mass $M$, renormalization scale $\mu$   
and effective threshold $s_0^{H^{(*)}}$  -- generally depends on 
the quantum numbers of quark currents  in the correlation function. 
In our analysis, for all 
mesons containing one and the same heavy quark, $b$ or $c$, 
a uniform range of $M^2$ and $\mu $ is chosen, whereas
the effective threshold depends also on the  light-quark flavour and 
spin-parity of the interpolating quark currents.

In  the heavy-quark limit  of the sum rules (see sect.\,\ref{sect:HQL} below), one expects
that an ``optimal  Borel window'' discussed in the previous section 
is located around $M^2\sim 2 m_Q\tau$, where 
$\tau\sim 1 \,\GeV \gg \Lambda_{QCD}$ does not   
scale with the heavy mass. Specifically,
for the correlation functions with $b$ quark ($c$ quark)  
we adopt the lower boundary of the Borel parameter $M_{min}^2= 4.5 ~\GeV^2$ 
($M_{min}^2= 1.5~\GeV^2$), so that 
the magnitude of the sum over $d=4,5,6$ condensate contributions  
to the OPE 
does not exceed  $\pm 5\%$ of the perturbative part.
The renormalization scale $\mu$, being generally in the ballpark of $M$,
is chosen so that one retains 
the hierarchy of NNLO and NLO terms in the perturbative part 
of OPE. We fix $\mu=3.0 $ GeV ($\mu=1.5$ 
GeV) as ``default'' renormalization scale in the correlation functions 
with $b$  ($c$) quarks. With this choice, 
at $M^2\geq M^2_{min}$, the NNLO terms in OPE with $b$ quark ($c$-quark) 
are less than 15 \% (30 \%) of NLO terms. The scale dependence will be investigated by varying $\mu$ within the intervals indicated in Table ~\ref{tab:uncert}.
%%%%%%
\begin{table}[t]
{\small
%\scriptsize
\begin{center}
\begin{tabular}{|c||c|c|c|c|c|c|c|c|}
\hline
   Corr. function
& \multicolumn{2}{c|}{$\Pi_T^b(M^2)$}
& \multicolumn{2}{c|}{$\Pi_5^b(M^2)$}
& \multicolumn{2}{c|}{$\Pi_T^c(M^2)$}
& \multicolumn{2}{c|}{$\Pi_5^c(M^2)$}
\\
\hline
\hline
   default $M^2$
%Borel parameter
[GeV$^2$]
& \multicolumn{4}{c|}{5.5}
& \multicolumn{4}{c|}{2.0}
\\[-2mm]
%\hline
   (range) %in GeV$^2$
& \multicolumn{4}{c|}{(4.5\,--\,6.5)}
& \multicolumn{4}{c|}{(1.5\,--\,2.5)}
\\
\hline
   default $\mu$
%renorm. scale
[GeV]
& \multicolumn{4}{c|}{3.0}
& \multicolumn{4}{c|}{1.5}
\\[-2mm]
   (range)
& \multicolumn{4}{c|}{(3.0\,--\,5.0)}
& \multicolumn{4}{c|}{(1.3\,--\,3.0)}
\\
\hline
%  $r^{\rm pert, NNLO}(M^2)$
%& -4.2\%
%& -2.1\%
%& 14.0\%
%& 14.0\%
%& -1.6\%
%& -1.1\%
%& 30.1\%
%& 26.9\%
%\\
%\hline
%   $r^{\text{(d$\geq$4)}}(M^2)$
%& -5.2\%
%& -2.6\%
%& -1.1\%
%& -0.5\%
%& -5.2\%
%& -2.6\%
%& 3.7\%
%& 2.5\%
%\\
\hline
   Sum rule for
& $f_{B^*}$
& $f_{B^*_s}$
& $f_{B}$
& $f_{B_s}$
& $f_{D^*}$
& $f_{D^*_s}$
& $f_{D}$
& $f_{D_s}$
\\
\hline
   Meson mass [GeV] \cite{PDG}
& 5.325
& 5.415
& 5.280
& 5.367
& 2.010
& 2.112
& 1.870
& 1.968
\\
\hline
  eff. threshold [GeV$^2$]
& 34.1
& 36.3
& 33.9
& 35.6
& 6.2
& 7.4
& 5.6
& 6.3
\\
\hline
central value [MeV] & $210.3$& $251.4$& $206.7$& 
$241.7$&$241.9$&$293.3$& $201.0$&$237.4$\\ \hline
%&&&&&&&&\\
$\Delta M^2$&
$^{+0.1}_{-1.8}$& $^{+3.8}_{-5.4}$&$^{+6.1}_{-4.5}$&$^{+8.1}_{-5.8}$&
$^{+3.6}_{-5.0}$& $^{+10.1}_{-9.5}$& $^{+10.7}_{-12.1}$ & 
$^{+8.6}_{-19.4}$\\ \hline
%%%%%%%%
$\Delta \mu$& $^{+0.0}_{-5.3}$&$^{+0.0}_{-8.6}$& 
$^{+13.0}_{-0.0}$&$^{+10.3}_{-0.0}$& $^{+17.3}_{-3.9}$&  
$^{+12.3}_{-2.3}$&$^{+1.3}_{-3.5}$& $^{+3.5}_{-9.3}$\\\hline
%%%%%%%
$\Delta m_Q$&$^{+9.0}_{-8.7}$&$^{+9.9}_{-9.7}$&$^{+7.6}_{-7.5}$&$\pm 8.2$&
  $\pm7.5$& $\pm8.0$&$^{+1.6}_{-1.9}$& $^{+1.7}_{-2.1}$\\
\hline
%%%%%%%
$\Delta m_s$& --& $\pm 1.5$& -- &$\pm 1.6$ &--& $\pm2.2$&--& 
$\pm 3.1$\\ \hline
%%%%%
$\Delta \langle q\bar q\rangle$ & $\pm 3.2$& $^{+2.3}_{-2.4}$ 
&$^{+2.8}_{-2.9}$& $^{+2.1}_{-2.2}$ &$\pm 4.0$& 
$^{+2.8}_{-2.9}$&$\pm3.0$&$\pm2.2$\\ \hline
%$~$\Delta \langle s\bar s\rangle$
%%%%%%%
$\Delta \frac{\langle s\bar s\rangle}{\langle q\bar 
q\rangle}$&--&$^{+7.1}_{-7.3}$&--&$^{+6.5}_{-6.7}$&--& 
$^{+8.6}_{-8.9}$&--&$^{+6.6}_{-6.9}$\\ \hline
%%%%%
$\Delta \langle GG\rangle$&$^{+0.4}_{-0.2}$& 
$^{+0.4}_{-0.2}$&$^{+0.1}_{-0.3}$& $^{+0.1}_{-0.3}$& 
$^{+1.9}_{-0.9}$&$^{+1.7}_{-0.9}$& $^{+0.4}_{-0.8}$& 
$^{+0.4}_{-0.8}$\\\hline
%%%%
$\Delta m_0^2$& $\pm 0.9$& $\pm0.7$& $\pm0.3$&$\pm 0.2$ 
&$\pm0.7$&$\pm0.5$& $\pm0.5$& $\pm0.4$\\\hline
%%%%%
$\Delta d456$& $\pm 4.0$& $\pm3.2$& $\pm 0.9$& $\pm 0.6$& $\pm4.6$& 
$\pm3.8$&$\pm2.8$&$\pm2.5$\\
\hline
\end{tabular}
\end{center}
\caption{ \it  Details of the numerical analysis of QCD sum rules
for decay constants. $\Delta M^2$ etc. denote the individual  
uncertainty of
decay constant (in MeV) due to the variation of $M^2$ etc.  within the
adopted interval.}
\label{tab:uncert}
}
\end{table}

%%%%%%%%%
Finally, in the sum rule for each decay constant $f_{H^{(*)}}$ 
we determine the effective threshold $s_0^{H^{(*)}}$, demanding
that the measured mass of the $H^{(*)}$-meson 
is reproduced from the differentiated  sum rule with an accuracy no less than 0.5\%.  
Although we work in the isospin symmetry limit, in the  numerical
analysis, for definiteness, mesons  with the flavour content 
$c\bar d$ and  $b\bar d$ are taken.
The Borel parameter is constrained from above, $M^2<M_{max}^2$,
so that the relative contribution of excited and continuum states 
to the sum rule, which in our approximation is equal to 
$1- \Pi_{T(5)}^{(pert)}(M^2,s_0)/\Pi_{T(5)}^{(pert)}(M^2,\infty)$, remains less than $50 \%$. 

The decay constants  are numerically calculated from the 
sum rules at $M^2_{min}<M^2<M^2_{max} $, 
taking for each $M^2$ the fitted value of $s^{H^{(*)}}_0$\!. The results
obtained at central input are presented in Table~\ref{tab:uncert}.
To estimate the total uncertainty 
of the calculated decay constants, we vary each input parameter separately, 
assuming (conservatively) that they are uncorrelated. 
Note on the other hand that  the values of the effective threshold $s^{H^{(*)}}_0$, 
after fixing the meson mass, 
are correlated with the Borel parameter. Hence, we do not attribute 
a separate uncertainty related to the choice of $s^{H^{(*)}}_0$. 
To take into account, albeit rather conservatively, 
the neglected $d>6$ terms in the OPE, we attribute to each 
calculated value of the decay constant 
an additional theoretical error  equal  to the  sum of  
$d=4,5,6$ condensate contribution.
All individual uncertainties -- except very small ones, with a magnitude
$\leq 0.1$ MeV --
are presented in Table~\ref{tab:uncert}. Adding them in quadrature,
we arrive at the final results for the decay constants of heavy-light 
vector and pseudoscalar mesons:

\ba
&f_{B^*}=(210^{+10}_{-12})\,[261]\,  \MeV\,,
&f_B~=(207^{+17}_{-09})\,[258]\, \MeV\,,
\label{eq:fBres}\\
&f_{B_s^*}=(251^{+14}_{-16})\,[296]\,\MeV\,,
&f_{B_s}=(242^{+17}_{-12})\,[285]\,\MeV\,,
\label{eq:fBsres}\\
&f_{D^*}=(242^{+20}_{-12})\,[297]\,\MeV\,,
&f_D~=(201^{+12}_{-13})\,[237]\,\MeV\,,
\label{eq:fDres}\\
&f_{D_s^*}=(293^{+19}_{-14})\,[347]\,\MeV\,,
&f_{D_s}=(238^{+13}_{-23})\,[266]\,\MeV\,.
\label{eq:fDsres}
\ea
where  the duality-independent upper bounds are presented in square brackets,
with the uncertainties included in the same way as in \cite{AKbounds}. 
Note that for charmed mesons the upper bounds are more restrictive than for bottom 
mesons.

As we can see from the above results, the QCD sum rule predictions 
for decay constants of vector and pseudoscalar have  similar uncertainties.
We also calculated the ratios of vector and pseudoscalar meson decay
constants dividing the corresponding  sum rules by each other:
\ba
&f_{B^*}/f_B= 1.02^{+0.02}_{-0.09}\,,
& f_{B_{s}^*}/f_{B_s}= 1.04^{+0.01}_{-0.08}\,,\label{eq:SRratiosB}\\
&f_{D^*}/f_D= 1.20^{+0.13}_{-0.07}\,,
& f_{D_{s}^*}/f_{D_s}=1.24^{+0.13}_{-0.05}\,.
\label{eq:SRratiosD}
\ea
The individual uncertainties for the above ratios  
are treated in the same way as for separate sum rules; in this 
case the correlations result in somewhat smaller  total uncertainties. 
Finally, we also obtain  the $SU(3)_{fl}$ violating ratios of
decay constants:
\ba
&f_{B_s}/f_B= 1.17^{+0.03}_{-0.04}\,,
& f_{B_{s}^*}/f_{B^*}= 1.20\pm0.04\, ,\\
&f_{D_s}/f_D= 1.18^{+0.04}_{-0.05}\,,
& f_{D_{s}^*}/f_{D^*}=1.21\pm0.05\,.
\label{eq:SU3ratios}
\ea

\section{ Other versions of  sum rules }

Apart from the uncertainties caused by the input parameters, 
the overall accuracy of QCD sum rules is influenced 
by the  quark-hadron duality approximation (\ref{eq:dual}).
One can argue that  the ``semi-local'' duality used here can be trusted, having in mind   
the positivity of the spectral function and the fact 
that the mass of the  ground-state hadron is reproduced 
from the differentiated sum rule with a high precision.
  
One possible strategy to assess the ``systematic'' uncertainty
of the adopted calculational procedure  
is to employ other  versions of QCD sum rules,
based on the same correlation function and same OPE, but
differing from the standard Borel sum rules by the 
weight function  multiplying the spectral density in the  
dispersion integrals.  In the standard version, after Borel 
transformation, the role of the weight function 
is played by the exponent $\exp(-s/M^2)$.  Transforming the 
initial, $q^2$-dependent dispersion integrals differently, 
one modifies the weight function resulting in a redistribution of the spectral density between the ground-state hadron and excited states.
Below  we consider a few different versions  of QCD sum rules for 
decay constants and carry out their numerical analysis.

\subsection*{\small A. Power moments}

First, we employ the well familiar power moments of 
the QCD sum rules \cite{SVZ}, obtained by differentiating 
over $q^2$ 
the hadronic dispersion relations (\ref{eq:dispPiT})  for the invariant amplitude 
at some spacelike value $q^2_0\ll m_Q^2$.
Minimum two differentiations are needed to get rid of 
subtraction terms. For the l.h.s.  of (\ref{eq:dispPiT}) we use the OPE  
result (\ref{eq:OPE}) and for the hadronic spectral
density of excited and continuum states the  quark-hadron duality 
approximation (\ref{eq:dual}). 
The decay constants squared are then determined as: 
\ba
 f_{H^*}^2= \frac{(m_{H^*}^2-q_0^2)^{n+1}}{m_{H^*}^2}\Pi_{T}^{(n)}(s_0^{H^*},q_0^2)\,,
~~ f_{H}^2= \frac{(m_H^2-q_0^2)^{n+1}}{m_H^4}\Pi_{5}^{(n)}(s_0^{H},q_0^2)\,,
\label{eq:Hilb}
\ea
where  the $n$-th  moment of the sum rule is: 
\ba
\Pi_{T(5)}^{(n)}(s_0,q_0^2)\equiv 
\int\limits_{(m_Q+m_q)^2}^{s_0}\frac{ds}{(s-q_0^2)^{n+1}}~\rho^{(pert)}_{T(5)}(s)
\nonumber\\
+
\left(\frac{d}{dq^2}\right)^n\Big[\Pi_{T(5)}^{\langle \bar{q}q\rangle}(q^2)
+\Pi_{T(5)}^{(d456)}(q^2)\Big]\Bigg|_{q^2=q_0^2}\,,
\label{eq:powmom}
\ea
with a power weight function in the dispersion integral. 
In the numerical analysis we employ only the $n=2,3$ moments to avoid
the growth of higher-dimensional condensate terms at larger $n$.
The effective threshold is again estimated by forming the 
ratio of the second and third moments and fitting the 
ground-state meson mass.
For bottom mesons the power moments of sum rules  work well
at $q_0^2=0$,  with both NNLO and $d\geq 4$ corrections sufficiently small. To fulfill the same criteria for charmed mesons,  
it is necessary to take $q_0^2 < 0$.  
The results for the decay constants obtained from 
the power moments are collected in Table~\ref{tab:altern}.
%%%%%%%%%%%%%%
%%%%%%%%%%%%%
\begin{table}[t]
{\small
\begin{center}
\begin{tabular}{|c||c|c|c|c||c|c|c|c|}
\hline
 Method& \multicolumn{8}{c|}{Decay constant [MeV]}
\\
\cline{2-9}
&$f_{B^*}$& $f_{B^*_s}$& $f_{B}$& $f_{B_s}$& $f_{D^*}$& $f_{D^*_s}$& 
$f_{D}$& $f_{D_s}$\\
\hline
power moments &
196 &236 & 198& 231 &228 & 281&203&238\\
\hline
Borel SR with $1/s$ weight &
  -- & -- & 211& 248 & -- & -- & 220 & 260
\\
\hline
Borel SR with $s$ weight&
  208 & 245 & 201 &  233 & 232 & 271 & 175 & 207
\\
\hline
Borel SR  w/o radial excit.&
  208& 249 & 208& 242 & 243 & 290&204& 239
\\
\hline
standard Borel SR &
210 & 251&207 &242 &242 & 293&201&238\\
\hline
\end{tabular}
\end{center}
\caption{\it Decay constants calculated from different sum rules at 
central input. In the power moments $q_0^2=0$ ($q_0^2=-4.0$ {\rm GeV}\,$^2$) is taken 
for bottom (charmed) mesons.}
\label{tab:altern}
}
\end{table}
%%%%%%%%%%%%%%

\subsection*{ \small B. Modified Borel sum rule}

Let us consider a combination of invariant amplitudes:
\be
\Delta_{T(5)}(q^2) \equiv \frac{\Pi_{T(5)}(q^2)-\Pi_{T(5)}(0)}{q^2}\,.
\label{eq:deltPi}
\ee
Note that this expression is finite  at $q^2=0$. 
Approximating  the l.h.s. by OPE, the r.h.s. by the dispersion relations 
and performing the Borel transformation, we obtain:
\be
\Delta^{(OPE)}_{T}(M^2)= f_{H^*}^2e^{-m_{H^*}^2/M^2}+\int\limits_{(m_H+m_P)^2}^\infty ds\, \frac{\rho^h(s)}{s}  e^{-s/M^2}\,,
\label{eq:delPiM}
\ee
and a similar relation for the pseudoscalar-meson channel.
An additional $1/s$  factor appears in the weight function 
multiplying the spectral density.
Applying the quark-hadron duality approximation  (\ref{eq:dual}),
we calculate the decay constants $f_{H^{(*)}}$ from these  sum rules.
Dividing (\ref{eq:delPiM}) by the conventional Borel sum rule 
we obtain a relation for the inverse mass squared of the ground-state meson,
allowing us to adjust the threshold  $s_0^{H^{(*)} }$.

It is even simpler to obtain another modified Borel sum rule with an extra
power of $s$ in the integral. One needs to multiply 
both parts of (\ref{eq:fBstarSR}) and  (\ref{eq:fBSR}) by $e^{m_{H^*}^2/M^2}$ and 
$e^{m_{H}^2/M^2}$, respectively, and 
after that differentiate them over $-1/M^2$. In this case the effective threshold 
is estimated by dividing the modified sum rule by the standard Borel sum rule
and adjusting the result to the mass squared of $H^{(*)}$. 

The numerical results for the decay constants obtained from the 
modified Borel sum rules with $1/s$ and $s$ weights are given in  Table~\ref{tab:altern}. 
Here we compare the results of different sum rules obtained
at one and the same central input  (as specified in Tables~\ref{tab:inp} and~\ref{tab:uncert}).
However, it turns out that for the vector-meson decay constants calculated from $1/s$ sum rules, the central values of Borel parameter specified in Table~\ref{tab:uncert} are not suitable, hence the corresponding results are missing.  

\subsection*{ \small C. Excluding the first radial excitation}

Radial (i.e., same spin-parity) excitations of heavy-light mesons form the resonance part of the 
hadronic spectrum above the ground states in the  spectral densities 
(\ref{eq:rhoHstar}) and (\ref{eq:rhoH}). 
Including these resonances in the hadronic spectral density 
explicitly, one improves  the accuracy of the quark-hadron duality approximation. 
Following this strategy, we separate the first radial excitation $H^{*'}$ from the rest of hadronic
spectrum in the vector-meson channel, transforming the 
duality ansatz (\ref{eq:dual}) to the following form: 
\be
\rho^h_T(s)\theta(s-(m_H+m_P)^2)= m^2_{H^{*'}}f^2_{H^{*'}}\delta(s-m_{H^{*'}}^2)+\rho^{(pert)}_T (s) \theta(s-s^{H^{*'}}_0)\,,
\label{eq:dualrad}
\ee
where the total width of $H^{*'}$ is neglected for simplicity 
(it can be easily restored employing a Breit-Wigner ansatz) and $s_0^{H^{*'}}$ generally differs from the effective threshold in (\ref{eq:dual}). In the same way, the 
spectral density in the pseudoscalar channel is modified introducing the 
excited state $H'$. 

Currently, only limited experimental data on the radially excited charmed mesons
are available.  The resonances $D(2550)$ and $D(2600)$, observed 
in \cite{BaBarDprime} (see also \cite{PDG})
represent realistic candidates for the first radially excited $D'$ and $D^{*'}$ states, 
respectively.  The mass differences between these resonances and ground-states 
$D$ and $D^*$ are in the same ballpark as for
the light unflavoured mesons, cf. the mass difference 
between the first radial excitation  $\rho'=\rho(1450)$ and 
the ground-state  $\rho$ meson. Here we assume that 
the mass differences between the first excited 
and ground states for all heavy-light mesons 
are approximately the same:
\be
m_{B'}-m_B\simeq m_{D'}-m_D\simeq m_{D_s'}-m_{D_s}\,, 
~~  
m_{B^{*'}}-m_{B^*}\simeq m_{D^{*'}}-m_{D^*}
\simeq m_{D_s^{*'}}-m_{D_s^*}\,.
\label{eq:radmass}     
\ee

Without introducing extra parameters, such as the decay constants 
of the radially excited mesons, we suggest to use a modified QCD sum rule
in which the spectral density is multiplied by an additional factor
$(m_{H^{(*)'}}^2-s)$ vanishing at the position of the 
first radially excited state $H^{(*)'}$. To derive this sum rule, one simply 
multiplies the initial, $q^2$-dependent dispersion relation for $\Pi_{5(T)}(q^2)$
by an  overall factor $(m_{H^{(*)'}}^2-q^2)$. After Borel transformation
the following expression, e.g.,  for the decay constant of 
the vector heavy-light meson  is obtained:
\ba
f_{H^*}^2=\frac{e^{m_{H^*}^2/M^2}}{m_{H^*}^2(m_{H^{*'}}^2-m_{H^*}^2) }
\Bigg\{
\int\limits_{(m_Q+m_q)^2}^{s_0^{H^{*'}}}
\!ds (m_{H^{*'}}^2-s)\, e^{-s/M^2}\rho^{(pert)}_{T} (s)
\nonumber\\
 +
\Big (m_{H^{*'}}^2-\frac{d}{d(-1/M^2)}\Big)\big[\Pi_{T}^{\langle \bar{q} q \rangle} (M^2)+
\Pi_{T}^{(d456)} (M^2)\big]
\Bigg \}\,.
\label{eq:fBSR1}
\ea
It is straightforward to derive the analogous  sum rule for $f_{H}$.

We take as an input  $m_{D'}=2.55$ GeV , $ m_{D^{*'}}=2.60$ GeV
and estimate the masses of other radially excited states
from (\ref{eq:radmass}). In fact, in   the above sum rule 
we do not necessarily need a precise value of the mass $m_{H^{(*)'}}$. 
Important is that, due to a partial cancellation 
between the two intervals, below and above $s=m^2_{H^{(*)'}}$, the 
region above the ground state 
and adjacent to the first radial excitation
is suppressed in the integral over the weighted spectral density.
As a result, the r.h.s. of (\ref{eq:fBSR1}) becomes 
less sensitive to the duality approximation, allowing 
us to simplify the choice of the effective threshold.
Here we simply adopt $s^{H^{(*)'}}_0=m_{H^{(*)'}}^2$, without adjusting 
the threshold parameter. Interestingly, as our numerical analysis shows, 
this choice reproduces the  masses of the ground states from differentiated sum rules (\ref{eq:fBSR1}) within 1\% accuracy.
The decay constants obtained from (\ref{eq:fBSR1}) and from the
analogous sum rule for vector mesons are surprisingly close to the 
decay constants obtained from standard Borel sum rules with fitted
effective thresholds (see Table.~\ref{tab:altern}).

Assessing the mutual deviations between the decay constants 
calculated from various sum rules considered in this section, we have to include the variations of all entries in Table \ref{tab:altern} due to the input parameters. After that, we find that the predicted intervals of all decay constants calculated from different sum rules overlap.   
The same is valid  for the ratios of decay constants .

\section{Heavy-quark limit of the sum rules}\label{sect:HQL}

The  heavy-quark  mass expansion in QCD sum rules for decay constants 
was pioneered in \cite{Shuryak}. Later, the sum rule  technique was applied 
\cite{Neubert91,Braunetal,BroadGrozin} in the framework of  the heavy-quark effective theory (HQET), considering, instead of 
the quark currents with a finite $m_Q$,  their HQET counterparts, with a possibility 
to systematically resum the logarithms $\ln(m_Q/\mu)$ 
emerging in the OPE . 
The ``static'' value  of $f_{H^{(*)}}$  calculated from the HQET sum rule
in the $m_Q\to \infty$ limit, receives large inverse heavy-mass corrections
which have to be estimated separately \cite{Braunetal,Neubert_rev,PS}.

To obtain the heavy-quark  limit of the 
sum rules (\ref{eq:fBstarSR}) and  (\ref{eq:fBSR}), one has to rescale 
the $m_Q$-dependent  parameters:
\be
m_H=m_Q+\bar{\Lambda},~~
s_0^H=m_Q^2+2m_Q\omega_0,~~
M^2=2m_Q\tau\,.
\label{eq:hqlimit}
\ee
After this replacement, 
the sum rule for the heavy-light vector meson decay constant 
transforms to:
\begin{align}
 f_{H^*}^2m_{H^*}\Big(\frac{m_{H^*}}{m_Q}\Big)
e^{-\frac{\bar\Lambda}{\tau}-\frac{\bar\Lambda^2}{2m_Q\tau}}=&\frac{\tau^3}{\pi^2}
\int_{0}^{\frac{\omega_0}{\tau}}dz\,e^{-z}\Bigg(\frac{z^2}{1+\frac{2z\tau}{m
_Q}}\Bigg)\Bigg(2+\frac{1}{1+\frac{2z\tau}{m_Q}}\Bigg)
\nonumber \\
&\times\Bigg\{1+\frac{2\alpha_s}{\pi}\Bigg[\ln\Big(\frac{m_Q}{2\tau}\Big)+\frac{3}{2
}+\frac{2\pi^2}{9}-\ln(z)+\frac{2}{3}\,\mathcal{K}_T\Big(\frac{2z\tau}{m_Q}\Big)\Bigg]
\Bigg\}
\nonumber \\
&-\langle q\bar
q\rangle\Bigg\{1+\frac{2\alpha_s}{3\pi}\Bigg(3+\Big(\frac{2\tau}{m_Q}\Big)\int_
0^\infty dz\,\frac{e^{-z}}{(1+\frac{2z\tau}{m_Q})^2}\Bigg)\Bigg\}\nonumber \\
&-\frac{\langle GG\rangle}{12 m_Q}+\frac{m_0^2\langle q \bar
q\rangle}{16\tau^2}+\frac{\pi \alpha_s r_{\rm vac}\langle q \bar
q\rangle^2}{ 162\tau^3}\Bigg\{1-\frac{16\tau}{m_Q}-\frac{32 \tau^2}{m_Q^2}\Bigg\}\,,
\label{eq:HQBstar}
\end{align}
where 
\ba
\mathcal{K}_T(x)=2\,{\rm
Li}_2(-x)+\ln(x)\ln(1+x)+\frac{x}{(3+2x)}\ln(x)
\nonumber \\
+\frac{(1+2x)(2+x)(1+x)}{(3+2x)x^2}\ln(1+x)
+\frac{6x^2+3x-8}{4(3+2x)x}-\frac{9}{4}\,,
\ea
so that 
\be
\lim\limits_{x\to 0}  \mathcal{K}_T(x)= 
\frac{4}{3}x\ln(x)-\frac{29}{18}x+\mathcal{O}(x^2)\,.
\ee
Note that deriving (\ref{eq:HQBstar}) we use the pole-mass scheme for $m_Q$,
as it is more convenient for the matching with HQET. The 
expression for the $O(\alpha_s)$ correction was accordingly modified.
 
The rescaled sum rule in the pseudoscalar channel \cite{Neubert91} is: 
\begin{align}
 f_H^2m_H\Big(\frac{m_H}{m_Q}\Big)^3
e^{-\frac{\bar\Lambda}{\tau}-\frac{\bar\Lambda^2}{2m_Q\tau}}=&\frac{3\tau^3}{\pi^2}
\int_{0}^{\frac{\omega_0}{\tau}}dz\,e^{-z}\Bigg(\frac{z^2}{1+\frac{2z\tau}{m
_Q}}\Bigg)
\nonumber\\
&\times \Bigg\{1+
\frac{2\alpha_s}{\pi}\Bigg[\ln\Big(\frac{m_Q}{2\tau}\Big)+\frac{1
3}{6}+\frac{2\pi^2}{9}-\ln(z)+\frac{2}{3}\,\mathcal{K}_5\Big(\frac{2z\tau}{m_Q}\Big)\Bigg]\Bigg\}
\nonumber\\
&-\langle
q\bar q\rangle\Bigg\{1-\frac{2\alpha_s}{3\pi}\Bigg(-1+3\frac{2\tau}{m_Q}\int_0^\infty
dz\,\frac{e^{-z}}{1+\frac{2z\tau}{m_Q}}\Bigg)\Bigg\}+\frac{\langle GG\rangle}{12 m_Q} 
\nonumber\\
&+\frac{m_0^2\langle q \bar
q\rangle}{16\tau^2}\Bigg\{1-\frac{4\tau}{m_Q}\Bigg\}+\frac{\pi \alpha_s
r_{\rm vac} \langle q \bar q\rangle^2}{162\tau^3}\Bigg\{1+\frac{6\tau}{m_Q}-\frac{48
\tau^2}{m_Q^2}\Bigg\}\,,
\label{eq:HQB}
\end{align}
where 
\be
\mathcal{K}_5(x)=2\,{\rm
Li}_2(-x)+\ln(x)\ln(1+x)-\frac{x}{1+x}\ln(x)+\frac{1+x}{x}\ln(1+x)-1\,,\\
\ee
and
\be
\lim\limits_{x\to 0}  \mathcal{K}_5(x)=-\frac{3}{2}x+\mathcal{O}(x^2)\,.
\ee

It is now possible to take the limit $m_Q\to \infty$ 
in (\ref{eq:HQBstar}) and (\ref{eq:HQB}), whereas the logarithms have to 
be treated separately. Neglecting the gluon radiative corrections, 
one reproduces the well-known heavy-quark limit of decay constants:
\be
f_H=f_{H^*}=\frac{\hat{f}}{\sqrt{m_H}}\,, 
\label{eq:HQL2}
\ee
where 
\be
\hat{f}= e^{\frac{\bar\Lambda}{2\tau}}\Bigg(\frac{3\tau^3}{\pi^2}
\int_{0}^{\frac{\omega_0}{\tau}}dz\,z^2e^{-z}
-\langle
q\bar q\rangle +
\frac{m_0^2\langle q \bar q\rangle}{16\tau^2}
+\frac{\pi \alpha_s
r_{\rm vac} \langle q \bar q\rangle^2}{162\tau^3}\Bigg)^{1/2}\,.
\label{eq:fhat}
\ee
Thus, the rescaled decay constant $\hat{f}$ receives contributions from the perturbative loop, quark condensate and the $d=5,6$ condensates. 
The  $d=4$  gluon condensate term enters the sum rules at the $1/m_Q$ level, 
together with other inverse mass corrections.

The radiative correction to the ratio of decay constants 
obtained from (\ref{eq:HQBstar}) and (\ref{eq:HQB})
in the heavy quark limit:
\be
\frac{f_{H^*}}{f_{H}}=1-\frac{2\alpha_s}{3\pi}, 
\label{eq:HstarH}
\ee
is in accordance 
with the well known $O(\alpha_s)$ correction to the 
heavy-quark spin symmetry relation which follows from the matching of HQET and full QCD heavy-light currents.

It is interesting to compare our predictions for the ratios of decay constants 
(\ref{eq:SRratiosB}) and 
(\ref{eq:SRratiosD}), obtained from QCD sum rules 
with finite quark masses,  with the HQET relation \cite{Neubert_rev}:
\be
\frac{f_{H^*}}{f_{H}}=\Big(1-\frac{2\alpha_s(m_Q)}{3\pi}\Big)
\Big[1+ \delta/m_Q\Big]\,,
\label{eq:HQSS}
\ee
where we introduce a short-hand notation for the combination of HQET
parameters determining the inverse mass correction.
For the values of pole masses we use $m_b^{pole}=4.6$ GeV   and 
$m_c^{pole}=1.5$  GeV which correspond (with $O(\alpha_s)$ accuracy) 
to the $\MSbar$ quark masses  in Table~\ref{tab:inp}. For bottom 
mesons the interval for the l.h.s.  of  (\ref{eq:HQSS}) taken
from  (\ref{eq:SRratiosB}) corresponds to $\delta=180\div 650$ MeV.
Assuming the same value of this parameter for charmed mesons and neglecting $O(1/m_Q^2)$ corrections,
we obtain from (\ref{eq:HQSS}) the ratio $f_{D^*}/f_D=1.03\div 1.33$,
which agrees with our predicted interval  in (\ref{eq:SRratiosD}).
Note that the HQET parameters  contributing to $\delta$ were 
estimated from sum rules in HQET \cite{Neubert_rev} yielding 
$f_{B^*}/f_B=1.07\pm 0.02 $  and $f_{D^*}/f_D=1.35\pm 0.05 $
which is in a satisfactory agreement with our results from full QCD sum rules.
Numerically, the heavy-quark spin symmetry for decay constants is violated by 
an inverse mass correction of about 12-14\% (20-40\%) for bottom (charmed)
mesons. Let us finally estimate the heavy-quark flavour symmetry violation. E.g., the leading-order
ratio in HQET including radiative corrections \cite{Neubert_rev}:
\be
\frac{f_B}{f_D}=\sqrt{\frac{m_D}{m_B}}\Big(\frac{\alpha_s(m_c)}{\alpha_s(m_b)}
\Big)^{6/25}\Big(1+0.894\frac{\alpha_s(m_c)-{\alpha_s(m_b)}}{\pi}\Big)\simeq 0.69\,,
\label{eq:fBfD}
\ee
has to be compared with the interval $f_B/f_D\simeq 0.93\div1.19$ 
allowed by the intervals of sum rule predictions presented in (\ref{eq:fBres})
and (\ref{eq:fDres}), assuming no correlation between uncertainties.

\section{Discussion}

In this paper we calculated the  decay constants 
of heavy-light vector and pseudoscalar mesons employing 
the well established method  of QCD sum rules. The sum rules
for $B^*_{(s)}$ and $D^*_{(s)}$ mesons have acquired the same level of accuracy 
as the sum rules for $B_{(s)}$ and $D_{(s)}$ mesons: $O(\alpha_s^2)$ in the perturbative part 
and $O(\alpha_s)$ in the quark condensate term.
We correspondingly updated the numerical values of all decay constants, together 
with their upper bounds  and ratios. 
The uncertainties for $f_B$ and $f_D$ caused by the input variation 
became somewhat smaller than in the earlier sum rule determinations where $f_B=210\pm 19$ MeV, $f_{B_s}=244\pm 21$ MeV \cite{JL} and $f_B=206\pm 20$ MeV, $f_D=195\pm 20$ MeV
\cite{PS} were obtained, with a typical error of about $\pm 20$ MeV. 
This improvement is mainly due to smaller uncertainties of  quark 
$\MSbar$ masses achieved in recent years.
In this paper we investigated different versions of sum rules: 
power moments and Borel sum rules with a  modified weight of the spectral density. Within uncertainties, their predictions agree with the ones
obtained from the standard Borel sum rules.

Reducing the uncertainties of quark masses  and condensate densities is one of the few
remaining possibilities to further improve the accuracy of QCD sum rules for decay constants.
It is also desirable to obtain a fully analytic form 
of the $O(\alpha_s^2)$ corrections, in order to achieve
a better control over the renormalization scale dependence. Calculating the $O(\alpha_s)$ correction to the 
quark-gluon condensate term can also be useful, at least for the ratio of sum rules
for vector and pseudoscalar mesons where this contribution is enhanced.

Turning to  the comparison with  recent sum rule determinations of
pseudoscalar meson decay constants,
let us note that,  contrary to the analysis presented
in  \cite{Narison_new}, we do not  attempt to fit the heavy quark mass simultaneously
with the decay constants. We also cannot confirm the total uncertainties
and upper bounds quoted in  \cite{Narison_new}, which are both systematically
smaller than what is obtained here.
In~\cite{Melikhov},  while determining the pseudoscalar heavy-light meson decay constants
with $O(\alpha_s^2)$ accuracy,
an explicit polynomial  dependence of the effective threshold
on the Borel parameter is introduced.  The claim
that this dependence improves the sum rules and allows
one to estimate a related systematic error, remains obscure to us.
A  determination of  $f_{B_{(s)}}$  and $f_{D_{(s)}}$  from finite-energy sum rules
\cite{Schilcher}, which is a different method based on the same correlation function and
OPE,  yields somewhat smaller decay constants than the ones obtained here.

In Table \ref{tab:compar} we compare our predictions for 
decay constants and their ratios with the most recent lattice QCD determinations,  
revealing  a good  agreement within the uncertainties.
  %%%%%%%%%%%
\begin{table}[h]
\begin{center}
\begin{tabular}{|c|c|c|}
\hline
 Decay constant & Lattice QCD [ref.] & this work\\
\hline
& 196.9 $\pm$ 8.9 \cite{Fermilab}& \\[-3mm]
$f_{B}$[MeV] & & $207^{+17}_{-09}$\\[-3mm]
& 186.0 $\pm$ 4.0 \cite{HPQCDb} & \\
\hline
& 242.0 $\pm$ 9.5 \cite{Fermilab} &\\[-3mm]
$f_{B_s}$[MeV] & &$242^{+17}_{-12}$\\[-3mm]
& 224 $\pm $ 5.0 \cite{HPQCDb} &\\
\hline
&1.229$\pm$ 0.026 \cite{Fermilab}   &\\[-3mm]
$f_{B_s}/f_B$ & & $1.17^{+0.03}_{-0.04}$\\[-3mm]
&1.205$\pm$ 0.007 \cite{HPQCDb}   &\\
\hline\hline
& 218.9 $\pm$ 11.3 \cite{Fermilab} &\\[-3mm]
$f_D$[MeV] & &$201^{+12}_{-13}$\\[-3mm]
& 213 $\pm$ 4.0  \cite{HPQCDc} &\\
\hline
& 260.1 $\pm$ 10.8 \cite{Fermilab} &\\[-3mm]
$f_{D_s}$[MeV]& &$238^{+13}_{-23}$\\[-3mm]
 &248.0 $\pm$ 2.5 \cite{HPQCDc} &\\
\hline
&1.188$\pm$ 0.025  \cite{Fermilab}  &\\[-3mm]
$f_{D_s}/f_D$  &&$1.18^{+0.04}_{-0.05}$\\[-3mm]
&1.164$\pm$ 0.018 \cite{HPQCDc}&\\
\hline\hline
&&\\[-4mm]
$f_{D^*}$[MeV] &278 $\pm$ 13 $\pm$ 10 \cite{fDstarlat}&$242^{+20}_{-12}$\\[2mm]
\hline
&&\\[-4mm]
$f_{D^*_s}$[MeV] &311 $\pm$ 9.0 \cite{fDstarlat} &$293^{+19}_{-14}$\\[2mm]
\hline
&&\\[-4mm]
$f_{D^*_s}/f_{D^*}$  &1.16$\pm$ 0.02$\pm$ 0.06 \cite{fDstarlat}   &$1.21\pm0.05$\\[2mm]
\hline
\end{tabular}
\end{center}
\caption{\it Decay constants of heavy-light mesons, comparison 
with lattice QCD  results.}
\label{tab:compar}
\end{table}
%%%%%%%%%%%

We can also compare the predictions for $f_D$ and $f_{D_s}$ 
with the averages \cite{RosnStone} over various experiments measuring 
$D_{(s)}\to \ell \nu_\ell$ decay widths:
$f_D^{(exp.av.)}=206.7\pm 8.5\pm 2.5$ MeV and $f_{D_s}^{(exp.av.)}=260.0 \pm 5.4$ MeV. 
We notice some  tension of our prediction 
(and also of the lattice result \cite{HPQCDc})  for $f_{D_s}$  with  the  above interval.
Furthermore, the most recent measurement of $B\to \tau\nu_\tau $ 
\cite{BelleBtaunu} yields $f_B^{(exp)}=(211 \pm 22 \pm 14)\,\text{MeV}/(|V_{ub}|/0.0035)$ 
taking for $|V_{ub}|$ a typical value obtained \cite{PDG,KMOW} 
from exclusive semileptonic $B$ decays.
A future reduction of the experimental uncertainty in this measurement  opens up
a possibility to use $f_B$ calculated in QCD  for an independent  $|V_{ub}|$ -determination.
On the other hand, the $|V_{ub}|$ independent ratio of $B\to \pi\ell\nu_e$   
and $B\to \tau \nu_\tau$ widths,  can be used \cite{KMOW} to check 
QCD calculations of form factors and decay constants.

We conclude our discussion mentioning the role 
of radial excitations of heavy-light mesons in the sum rules.
As we found, excluding the first radial excitation from the hadronic spectrum 
makes the sum rule less sensitive to the value of the effective threshold.
One can turn this argument around, anticipating that the sum rules
considered in this paper
are also capable to yield estimates of the decay constants 
for the first radial excitations of heavy-light mesons.
We plan a separate study in this direction.  

\newpage
\section{Acknowledgments}
We acknowledge useful discussions with Bj\"orn O. Lange.
This work is supported by  DFG Research Unit FOR 1873 ``Quark Flavour Physics
and Effective Theories'',  Contract No.~KH 205/2-1. 
AAP acknowledges partial support from RFFI  grant 11-01-00182-a.

\appendix
\section*{Appendix: OPE expressions}

\section{Perturbative spectral density}\label{App:perturbative}

Here we collect the expressions for NLO,  $O(\alpha_s)$ contributions to the
spectral density $\rho^{(\rm pert)}_{T(5)}(s)$ in the $\overline{\rm 
MS}$-scheme for the heavy quark mass $m_Q$.
For the vector heavy-light quark currents, according to our convention 
for the invariant
amplitude $\Pi_T(q^2)$, we extract the coefficient at $-g_{\mu\nu}$.
The corresponding spectral density reads:
\begin{align}
\rho_{T}^{(\rm pert,NLO)}(s)&=\frac{3C_F}{16\pi^2}s
\Bigg[1-\frac{5}{2}z+\frac{2}{3}z^2+\frac{5}{6}z^3+\frac{1}{3}z(-5-4z+5z^2)\ln(z)\nn\\
&-\frac{1}{3}(1-z)^2(4+5z)\ln(1-z)+\frac{2}{3}(1-z)^2(2+z)\Big(2\,{\rm{Li}}_2(z)\nn\\
&+\ln(z) \ln(1-z)\Big)-z(1-z^2)\left(3\ln\left(\frac{\mu^2}{m_Q^2}\right)+4\right)\Bigg]\,,
\label{eq:pertNLOT}
\end{align}
where $z=m_Q^2/s$ and ${\rm{Li}}_2(z)=-\int_0^z\frac{\ln(1-t)}{t}\,dt$.
%%%%%%%%
For the pseudoscalar heavy-light quark currents, one has 
\begin{align}
\rho_{5}^{(\rm pert,NLO)}(s)&=\frac{3C_F}{16\pi^2}(m_Q+m_q)^2 s (1-z)\Bigg[\frac{9}{2}(1-z)+(3-z)(1-2z)\ln(z)\nn\\
&-(1-z)(5-2z)\ln(1-z)+2(1-z)(2\,{\rm 
Li}_2(z)+\ln(z)\ln(1-z))\nn\\
&+(1-3z)\left(3\ln\left(\frac{\mu^2}{m_Q^2}\right)+4\right)\Bigg]\,.
\label{eq:pertNLO5}
\end{align}
%%%%%%%

For NNLO corrections we are using the results from \cite{ChetS}
calculated in the pole mass scheme. Hence, to properly apply  the 
$\overline{\rm MS}$
scheme for $m_Q$ to $\alpha_s^2$ accuracy, we have to add to  the NNLO part
the corrections which arise from expanding the pole mass in the  LO and NLO
in terms  of  $\overline{\rm MS}$  mass. For the vector-current 
correlation function they are
\begin{align}
\Delta_1\rho_{T}^{(\rm pert,NNLO)}(s)&=-\frac{3}{8\pi^2}sz\Bigg[(3-7z^2)r_m^{(1)^2}-2(1-z^2)r_m^{(2)}\Bigg]\,,\\ 
\Delta_2\rho_{T}^{(\rm pert,NNLO)}(s)&=-\frac{1}{16\pi^2}C_F\,r_m^{(1)}s\Bigg[-z(1-z^2)\Big(24\,{\rm 
Li}_2(z)+12\ln(z)\ln(1-z)\Big)\nonumber\\
&-2z\left(9+6z-17z^2\right)\ln(z)+2(1-z)(-4+5z+17z^2)\ln(1-z)\nonumber\\
&-z(1-z)(17+15z)\Bigg] \,,
\end{align}
respectively, and for the pseudoscalar-current  correlation function:
\begin{align}
\Delta_1\rho_{5,\rm
LO}^{\rm(pert,NNLO)}(s)&=\frac{3(m_Q+m_q)^2}{8\pi^2}s\Bigg[(3-20z+21z^2)r_m^{(1)^2}-2(1-
z)(1-3z)r_m^{(2)}\Bigg]\,,\\
\Delta_2\rho_{5}^{\rm(pert,NLO)}(s)&=-\frac{3(m_Q+m_q)^2}{8\pi^2}C_F\,r_m^{(1)}s\Bigg[(1-z)(1-3z)\Big(4\,{\rm 
Li}_2(z)\nn\\
&+2\ln(z)\ln(1-z)\Big)+(3-22z+29z^2-8z^3)\ln(z)\nn\\
&-(1-z)(7-21z+8z^2)\ln(1-z)+\frac{1}{2}(1-z)(15-31z)\Bigg]\,,
\end{align}
where $r_m^{(1,2)}$ are the well-known coefficients in the perturbative 
relation between the pole and $\overline{\rm MS}$  quark masses (given, e.g. in  
eqs. (B5-B9)) in \cite{JL}.

The corrections  due to the nonzero light-quark mass, after 
expanding the complete answer in powers of  $m_q$ read:
%%%%
\begin{align}
\delta\rho_{T}^{\rm(pert,LO,m_q)}(s)&=\frac{3}{8\pi^2}m_q\Bigg[2m_Q(1-z)-m_q(1+z^2)\Bigg]\,,\\
\delta\rho_{T}^{\rm(pert,NLO,m_q)}(s)&=\frac{3}{8\pi^2}C_F m_qm_Q\Bigg[2(1-z)\left(2\,{\rm 
Li}_2(z)+\ln(z)\ln(1-z)\right)\nonumber\\
&+(3-4z-z^2)\ln(z)-(1-z)(5+z)\ln(1-z)+\frac{1}{2}(17-26z+z^2)\nonumber\\
&+3(1-2z)\ln\left(\frac{\mu^2}{m_Q^2}\right)\Bigg]\,.
\end{align}
The analogous corrections to the perturbative part of the
pseudoscalar-current correlation function are:
%%%%%%%%%
\begin{align}
\delta\rho_{5}^{\rm(pert,LO,m_q)}(s)&=\frac{3(m_Q+m_q)^2}{8\pi^2}\Bigg[2(1-z)m_Qm_q-2m_q^2\Bigg]\,,\\
\delta\rho_{5}^{\rm(pert,NLO,m_q)}(s)&=\frac{3(m_Q+m_q)^2}{8\pi^2}C_F\,m_Q
m_q\left[(1-z)\Big(4\,{\rm Li}_2(z)+2\ln(z)\ln(1-z)\right.\nn\\
&-2(4-z)\ln(1-z)\Big)+2(3-5z+z^2)\ln(z)\nn\\
&+2(7-9z)+3(2-3z)\ln\left(\frac{\mu^2}{m_Q^2}\right)\Bigg]\,.
\end{align}
We include the above corrections only for the $s$-quark and
up to the second (first) power  in $m_s$  in LO (NLO). We checked that 
the higher-power  corrections in $m_s$  are vanishingly small.

\section{Condensate contributions}\label{App:nonperturbative}
 
We present  the condensate contributions in two forms: with an explicit $q^2$ dependence 
(needed, e.g. for the power moments)  and after Borel transformation.

In the correlation function of vector currents, the total 
contribution of the quark condensate is:
\begin{align}
 \Pi^{\langle \bar q q\rangle}_{\mu\nu} (q^2)&= \langle \bar q q\rangle \frac{m_Q}{m_Q^2-q^2}\left[g_{\mu\nu} \left(1-\frac{m_qm_Q}{2(m_Q^2-q^2)}+\frac{\alpha_sC_F}{2\pi} f_{V,1}(z)\right)\right.\nn\\
&\phantom{ =\langle \bar q q\rangle \frac{m_Q}{m_Q^2-q^2}\Big[}\left.-\frac{q_\mu q_\nu}{q^2}\frac{\alpha_sC_F}{\pi}f_{V,2}(z)\right]. 
\label{eq:qqbarq2}
\end{align}
with the NLO  terms given by 
\begin{align}
 f_{V,1}(z) &= 2-z+z\left(1-z\right)L_z-\frac{z}{z-1}\left(3\ln\frac{\mu^2}{m^2}+4\right)\;,\\
 f_{V,2}(z) &= 1-2z+2z\left(1-z\right)L_z\;,
\end{align}
where the short-hand notations $z = \frac{m_Q^2}{q^2}$ and $L_z= \ln\left(\frac{z-1}{z}\right)$ are used. In the case $q=s$  the first-order $O(m_q)$  correction 
included in  (\ref{eq:qqbarq2}) provides a sufficient accuracy.
For our purpose, only the coefficient $\Pi^{\langle \bar q q\rangle}_T (q^2)$ of the structure  
$ -g_{\mu\nu}$ is needed.  The Borel-transformed  expression of this amplitude  is:
\begin{align}
 \Pi^{\langle \bar q q\rangle}_T (M^2)&=-m_Q\langle \bar q q\rangle e^{-\frac{m_Q^2}{M^2}} \left(1-\frac{m_qm_Q}{2M^2}+\frac{\alpha_sC_F}{2\pi}\left[1-3\frac{m_Q^2}{M^2}\ln\frac{\mu^2}{m_Q^2}-4\frac{m_Q^2}{M^2}\right.\right.\\\nonumber
 &\left.\left.+\frac{m_Q^2}{M^2}e^{\frac{m_Q^2}{M^2}}\Gamma\Big(-1,\frac{m_Q^2}{M^2}\Big)\right]\right)\;,
\end{align}
%%%
with the incomplete gamma function $\Gamma(a,z) = \int_z^\infty t^{a-1} e^{-t} dt$.
The NLO part in (\ref{eq:qqbarq2}) originating from one-loop  diagrams 
has an imaginary part at $q^2\to s \geq m_Q^2$ . The latter, in addition to the terms 
proportional to $\delta(s-m_Q^2)$ and its derivatives,  
contains also a  part  which does not vanish at $s> m_Q^2$, that is proportional 
to $\theta(s-m_Q^2)$.
Since we include the latter in the OPE spectral density involved in the quark-hadron 
duality approximation, we present here also the spectral density 
of the condensate contribution: 
\begin{align}
 \rho^{\langle \bar q q\rangle}_T(s)  &= -m_Q\langle \bar q q\rangle \Big(\delta(m_Q^2-s)
-\frac{1}{2}m_qm_Q\delta'(s-m_Q^2)
+\frac{\alpha_sC_F}{2\pi}\Big[\delta(m_Q^2-s)\\\nonumber
 &-m_Q^2\Big(3\ln\frac{\mu^2}{m_Q^2}+4\Big)\delta'(m_Q^2-s)+\frac{m_Q^2}{s^2}
\theta(s-m_Q^2)\Big]\Big)\;.
\end{align}
In the pseudoscalar-meson channel, the quark condensate contribution
to the correlation function  in the same approximation reads 
\begin{align}
\label{PSqq}
 \Pi^{\langle \bar q q\rangle}_5 (q^2) &= -\langle \bar q q\rangle\frac{(m_Q+m_q)^2m_Q}{m_Q^2-q^2}\left(1-\frac{m_q}{2m_Q}-\frac{m_qm_Q}{2(m_Q^2-q^2)}-\frac{\alpha_sC_F}{2\pi}
f_5(z) \right)\;,
\end{align}
 with the coefficient (see also \cite{JL})
\begin{align}
 f_5(z)= 3\frac{z}{z-1}\left(L_z\left(2-z\right)-1\right)+\frac{1}{z-1}\left(3\ln\frac{\mu^2}{m^2}+7-3L_z\right)\;.
\end{align}
 The Borel-transform of \eqref{PSqq} yields:
\begin{align}
 \Pi^{\langle \bar q q\rangle}_5 (M^2)&= -(m_Q+m_q)^2m_Q\langle \bar q q\rangle e^{-\frac{m_Q^2}{M^2}}\Bigg(1-\frac{m_q}{2m_Q}-\frac{m_qm_Q}{2M^2}
\\\nonumber &-\frac{\alpha_sC_F}{2\pi}\Big[\Big(3\ln\frac{\mu^2}{m_Q^2}+4\Big)\frac{m_Q^2}{M^2}-7-3\ln\frac{\mu^2}{m_Q^2}
 +3\Gamma\Big(0,\frac{m_Q^2}{M^2}\Big)e^{\frac{m_Q^2}{M^2}}\Big]\Bigg)\;.
\end{align}
The spectral density derived from \eqref{PSqq} reads
\begin{align}
 \rho^{\langle \bar q q\rangle}_5 (s)&= -(m_Q+m_q)^2m_Q\langle \bar q q\rangle \Bigg(\delta(m_Q^2-s)+\frac{\alpha_sC_F}{2\pi}\Big[\Big(7+3\ln\frac{\mu^2}{m_Q^2}\Big)\delta(m_Q^2-s)\\\nonumber 
 &-m_Q^2\Big(4+3\ln\frac{\mu^2}{m_Q^2}\Big)\delta'(m_Q^2-s)-\frac{3}{s} \theta(s-m_Q^2)\Big]\Bigg)\,.
\end{align}

%%%%%%%
The expressions for $d\geq 4$ condensate contributions 
for the vector-current correlation function read:
\begin{align}
 \Pi_{\mu\nu}^{\langle GG \rangle} (q^2) &= \frac{\langle GG \,\rangle}{12(m_Q^2-q^2)}  g_{\mu\nu}\,,~
\Pi_{\mu\nu}^{\langle \bar{q} G q \rangle} (q^2)=-\frac{m^2_0\langle \bar q q\rangle m_Q^3}{2(m_Q^2-q^2)^3} g_{\mu\nu}\;,\\
\Pi_{\mu\nu}^{\langle \bar{q} q \bar{q} q\rangle} (q^2) &=
\frac{8\pi  \alpha_s r_{vac}\langle\bar{q}q\rangle^2 }{81(m^2_Q-q^2)^4}\Bigg[\Big(9m_Q^4-16 m_Q^2q^2+4q^4\Big)g_{\mu\nu}
%\\\nonumber 
 % &
+\Big(10m_Q^2-4q^2\Big)q_\mu q_\nu\Bigg]\;.
\end{align}
The Borel-transformed form is:
\begin{align}
\Pi_{T}^{\langle GG \rangle} (M^2) &= - \frac{\langle GG \rangle }{12}  e^{-\frac{m_Q^2}{M^2}}\,,~~
\Pi_{T}^{\langle \bar{q} G q \rangle} (M^2)=\frac{ m^2_0\langle \bar q q \rangle m_Q^3}{4M^4}e^{-\frac{m_Q^2}{M^2}}\,,\\
\Pi_{T}^{\langle \bar{q} q \bar{q} q\rangle} (M^2)&= -\frac{32\pi\alpha_sr_{vac}\langle\bar{q} q\rangle^2}{81M^2} \Big(1+\frac{m_Q^2}{M^2}-\frac{m_Q^4}{8M^4}\Big) e^{-\frac{m_Q^2}{M^2}}\,.
\end{align}
The corresponding condensate contributions to the correlation function
with pseudoscalar currents are:
\begin{align}
 \Pi_{5}^{\langle GG \rangle} (q^2)&= \frac{\langle GG \rangle m_Q^2}{12(m_Q^2-q^2)} \,,~~
\Pi_{5}^{\langle \bar{q} G q \rangle} (q^2) = -\frac{m^2_0\langle \bar q q\rangle m_Q^3}{2(m_Q^2-q^2)^2}\Big(1-\frac{m_Q^2}{m_Q^2-q^2}\Big)\;,
\\
 \Pi_{5}^{\langle \bar{q} q \bar{q} q\rangle} (q^2) &= - \frac{8\pi \alpha_s r_{vac}
\langle \bar{q} q \rangle^2 m_Q^2q^2}{27(m_Q^2-q^2)^4}\Big(2q^2-3m_Q^2\Big)\;,
\end{align}
yielding after Borel transformation:
\begin{align}
\Pi_{5}^{\langle GG \rangle} (M^2)&= \frac{ \langle  GG\rangle m_Q^2}{12}  e^{-\frac{m_Q^2}{M^2}},~~
\Pi_{5}^{\langle \bar{q} G q \rangle} (M^2) =  -\frac{m^2_0\langle \bar q q\rangle m_Q^3}{2M^2}
\Big(1-\frac{m_Q^2}{2M^2}\Big) e^{-\frac{m_Q^2}{M^2}}\;,\\
\Pi_{5}^{\langle \bar{q} q \bar{q} q\rangle} (M^2) &= -\frac{16\pi\alpha_sr_{vac}\langle \bar{q} q \rangle^2m_Q^2}{27 M^2} \Big(1-\frac{m_Q^2}{4M^2}-\frac{m_Q^4}{12M^4}\Big) e^{-\frac{m_Q^2}{M^2}}\,.
\end{align}

%%%%%%%%%%%%%%%
\end{document}